\documentclass[12pt]{article}
\usepackage[mathscr]{eucal}
\usepackage{amssymb,latexsym}
\usepackage{verbatim}
\usepackage{amsmath}
\usepackage{amsthm}
\usepackage{enumerate}
\usepackage[utf8]{inputenc}
\usepackage{authblk}
\usepackage{color}
\usepackage{url}

\usepackage{tikz}
\usetikzlibrary{matrix}
\usetikzlibrary{snakes}
\usetikzlibrary{arrows,calc,shapes,decorations.pathreplacing}
\usepackage{tikz-cd}

\usepackage[normalem]{ulem}

\usepackage{amscd,amssymb,amsthm,verbatim}
\usepackage{enumerate}
\usepackage{bm}
\usepackage{mathrsfs, graphicx} 
\usepackage{stmaryrd}

\usepackage[%
bookmarks=true,
colorlinks,
linkcolor=black,
urlcolor=blue,
citecolor=blue,
plainpages=false,
pdfpagelabels,
final,
breaklinks=true\between
]{hyperref}
\usepackage{hyperref}

\newtheorem{thm}{Theorem}
\newtheorem{lem}[thm]{Lemma}
\newtheorem{Def}[thm]{Definition}
\newtheorem{prop}[thm]{Proposition}
\newtheorem{cor}[thm]{Corollary}

\newcommand\mc{\mathcal}

\renewcommand\l{\lambda}

\newcommand\bbR{{\mathbb R}}

\newcommand\wh{\widehat}

\newcommand\wt{\widetilde}

\renewcommand\S{\Sigma}

\newcommand\s{\sigma}

\newcommand\e{\epsilon}
\renewcommand\b{\beta}

\renewcommand\div{{\rm div}}

\newcommand\ric{{\rm Ric}}
\renewcommand\l{\lambda}
\newcommand\g{\gamma}

\renewcommand\a{\alpha}

\renewcommand\th{\theta}

\newcommand\beq{\begin{equation}}
\newcommand\eeq{\end{equation}}
\newcommand\ben{\begin{enumerate}}
\newcommand\een{\end{enumerate}}
\newcommand\bit{\begin{itemize}}
\newcommand\eit{\end{itemize}}

\DeclareMathOperator{\diver}{div}

\renewcommand{\div}{\diver}

\newcommand{\R}{\mathbb R}

\newcommand{\ov}{\overline}

\newcommand{\pd}{\partial}

\newcommand{\Z}{\mathbb{Z}}

\newcounter{mnotecount}

\title{The codimension 2 null cut locus with \\ applications to spacetime topology}

\author[1]{Gregory J. Galloway\footnote{galloway@math.miami.edu}}
\author[2]{Eric Ling\footnote{eling@math.rutgers.edu}}

\affil[1]{University of Miami, Coral Gables, FL}

\affil[2]{Rutgers University, New Brunswick, NJ}

\begin{document}
\date{}
\maketitle

\vspace{.15in}

\begin{abstract} 
In this paper we review and extend some results in the literature pertaining to spacetime topology while naturally utilizing properties of the codimension~2 null cut locus. 
Our results fall into two classes, depending on whether or not one assumes the presence of horizons.  Included among the spacetimes we consider are those that apply to the asymptotically (locally) AdS setting.

\end{abstract}


\tableofcontents

\newpage

\section{Introduction}

In this paper we review and extend some results in the literature pertaining to spacetime topology  from the point of view 
of the codimension 2 null cut locus, i.e. the null cut locus associated to a codimension 2 spacelike hypersurface in spacetime.  Our results fall into two classes, depending on whether or not one assumes the presence of horizons.  The results in which we do not assume the presence of horizons (as in \cite{Gannon, Lee, Costa, CostaMinguzzi, Stein}) may be interpreted as singularity results: Nontrivial topology (in a suitable sense) leads to future incompleteness.   The results in which we do assume the presence of horizons (as in \cite{GalBrowdy, LingLesourd}) pertain to the notion of topological censorship, which has to do with the idea that, outside of all black holes, one expects the topology to be simple (in suitable sense) at the fundamental group level.   

In \cite{Costa}, Costa e Silva presented a higher dimensional generalization of a well known result of Gannon \cite{Gannon} and Lee \cite{Lee}, in which the `enclosing surface' $S$ (within a given spacelike hypersurface) is not required to be simply connected.  We present a version of this result, in both the horizon and (as in \cite{Costa}) no horizon case.   An essential step in the proof is to construct a certain covering spacetime.  An issue arises in the covering construction in \cite{Costa} (see also \cite{Stein}) which we address here (see section \ref{gluing sec}).

An interesting effort in \cite{Costa}  was made
to lower the causality assumption from global hyperbolicity to causal simplicity. However, subsequently Costa e Silva and Minguzzi showed that causal simplicity does not necessarily lift 
to covers \cite{CostaMinguzzi}. To address this, they assume that the spacetime and its covers are past reflecting (a property weaker than causal simplicity).  As shown in \cite{CostaMinguzzi},  this assumption holds for spacetimes which admit a past complete conformal timelike Killing field. 

Our results apply to globally hyperbolic spacetimes, and also to a class of spacetimes that are not in general globally hyperbolic, but which admit an `exterior foliation' by globally hyperbolic spacetimes-with-time-like-boundary; see section \ref{TopologyNoHorizon} for details.  The latter spacetimes model spacetimes that are asymptotically locally anti-de Sitter, without having to introduce conformal completions.  
Our proofs involving these spacetimes make use of a reflectivity assumption in the `base', but do not require that this assumption lifts to covers.  As such we circumvent the difficulties mentioned in the previous paragraph.

Let $S$ be a closed separating hypersurface in a spacelike hypersurface $V$.  In Section~\ref{CutLocusProps} we study the null cut locus of $S$ with respect to the `inward' future directed null normal geodesics to $S$. In later applications we make a distinction between $S$-cut points and $S$-focal points.  While $S$-focal points lift to covers, $S$-cut points in general do not.   In later proofs we use the fact that $S$-focal points lift to covers, and that the existence of an $S$-focal point implies the existence of an $S$-cut point.  One can then apply properties of the null cut locus of $S$, as discussed in section~\ref{CutLocusProps}.  Apart from its analytic properties, 
the null cut locus has a natural relationship to achronal boundaries.
In sections \ref{TopologyNoHorizon} and \ref{TopologyHorizon} we discuss applications to spacetime topology in the no horizon and horizon case, respectively.

\section{The codimension 2 null cut locus and compactness results}\label{CutLocusProps}

\subsection{Null cut locus and the $s$ function}\label{cut locus section}

\medskip

Unless otherwise stated all objects are considered smooth.

Let $(M,g)$ be a spacetime and let $S$ be a codimension 2 spacelike submanifold. We define the \emph{future null tangent bundle} of $S$ as
\[
\mathcal{N}S^+ \,=\, \big\{v \in T_pM \mid p \in S \text{ and }  v \text{ is future directed null}\big\}.
\]
By our convention, the zero vector is not null  hence the zero section does not lie in $\mc{N}S^+$. We give $\mc{N}S^+$ the subspace topology inherited from the tangent bundle $TS$.

Define $s \colon \mc{N}S^+ \to [0, \infty]$ by 
\[
s(v) \,=\, \sup \{ t \geq 0 \mid d\big(S, \exp(tv)\big) = 0\},
\]
where $tv$ lies in the maximal domain of $\exp$, and where $d(S, \cdot) \colon J^+(S) \to [0, \infty]$ is the Lorentzian distance function restricted to $S$ defined via 
\[
d(S, p) \,=\, \sup\{L(\gamma) \mid \gamma \text{ is a future directed causal curve from $S$ to $p$}\}.
\]
Here $L$ is the Lorentzian arclength given by $L(\gamma) = \int \sqrt{-g(\gamma', \gamma')}$. If $v \in \mc{N}S^+$ and $\g(t) :=\exp(tv)$ extends to $[0, s(v)]$, then we say $\g\big(s(v)\big)$ is the \emph{future null cut point of $S$ along $\g$}. We will abbreviate this to \emph{$\g\big(s(v)\big)$ is the $S$-cut point along $\g$.} 

\smallskip

\begin{prop}\label{USC s prop}
Suppose $s(v) > 0$, and if $s(v) < \infty$ suppose $\g\big(s(v)\big)$ is the $S$-cut point along $\g$ where $\g(t) = \exp(tv)$. Then $s$ is upper semicontinuous at $v$. 
\end{prop}

\proof
$s$  is upper semicontinuous at $v$ provided $s(v) \geq \limsup_{v' \to v}s(v')$. This holds if and only if $s(v) \geq \limsup s(v_n)$ for all sequences $v_n \to v$. If $s(v) = \infty$, then we are done. Assume $s(v) < \infty$. Seeking a contradiction, suppose $v_n \to v$ and $s(v) < A:= \limsup s(v_n)$. By moving to a subsequence, we can assume $s(v_n) \to A$. Define $\gamma_n(t) = \exp (tv_n)$. Choose $\delta > 0$ such that $ b:= s(v) + \delta< A$ and $\gamma$ still extends to $[0, b]$. By moving to a subsequence, we can assume $\g_n$ is defined on $[0, b]$ as well. Since $b > s(v)$, there is a timelike curve from $S$ to $\g(b)$. Therefore $d\big(S, \g(b)\big) > 0$. However, by lower semicontinuity of the Lorentzian distance function \cite[Lem. 4.4]{BEE}, which holds for $d(S, \cdot)$ as well, we have 
\[
0 < d\big(S, \g(b)\big) \,\leq\, \liminf d\big(S, \g_n(b)\big) \,=\, \liminf 0 \,=\, 0, 
\]
which is a contradiction.
\qed

\medskip
\medskip

Note that this proposition does not require any causality assumptions.  It is used in the proofs of later results, in particular, in an essential way, in the proof of Lemma~\ref{W cpt lem horizon}.

The following proposition is given in \cite[Thm. 6.1]{Kemp}. While not actually needed in what follows, we include the proof for completeness, adding some relevant references.

\medskip

\begin{prop}\label{LSC s prop}
If $(M,g)$ is globally hyperbolic and $S$ is compact, then $s$ is lower semicontinuous.
\end{prop}

\proof
$s$ is lower semicontinuous at $v$ provided $s(v) \leq \liminf_{v' \to v}s(v')$. This holds if and only if $s(v) \leq \liminf s(v_n)$ for all sequences $v_n \to v$. This holds trivially for those sequences which $\liminf s(v_n) = \infty$. Therefore, seeking a contradiction, suppose $v_n \to v$ and $s(v) > A := \liminf s(v_n)$ with $A < \infty$. By moving to a subsequence, we can assume $s(v_n) \to A$. Fix $\delta > 0$ so that $b:= A + \delta < s(v)$. Define $b_n = s(v_n) + \delta$. Define $\g_n(t) = \exp (t v_n)$ and $\g(t) = \exp(tv)$. Since $\g$ is defined at $b$, by restricting to a subsequence, we can assume each $\g_n$ is defined at $b_n$. Set $p = \g(0)$, $q = \g(A + \delta)$, $p_n = \g_n(0)$, and $q_n = \g_n(b_n)$. Since $(M,g)$ is globally hyperbolic, a variation of \cite[Thm. 14.44]{ON} implies the existence of timelike geodesics $\s_n$ from $p_n$ to $q_n$ issuing orthogonally from $S$ which maximize the Lorentzian distance from $S$ to $q_n$. Since $S$ is compact, by restricting to a further subsequence, we can assume $p_n \to p' \in S$. By a limit curve argument \cite[Cor. 3.32]{BEE}, there is a causal curve $\s$ from $p'$ to $q$. 

Claim: $\s \neq \g|_{[0,b]}$. If this was not true, then the presence of $\s_n$ implies that the normal exponential map on $S$ would fail to be injective in a neighborhood of $\g|_{[0,b]}$. But then \cite[Prop. 10.30]{ON} implies that there must have existed a focal point of $S$ along $\g$ at or before $\g(b)$. Then \cite[Thm. 10.51]{ON} implies that there is a timelike curve from $S$ to $\g(b)$; hence $s(v) < b$, which is a contradiction. This proves the claim. Since $\s \neq \g|_{[0,b]}$, a `cutting the corner' argument within a normal neighborhood of $\g(b)$ implies $s(v) \leq A + \delta$, which is a contradiction.
\qed

\medskip

\begin{cor}\label{cont s cor}
Suppose $(M,g)$ is globally hyperbolic and for all $v \in \mc{N}S^+$ with $s(v) < \infty$ suppose that $\g\big(s(v)\big)$ is the $S$-cut point along $\g$ where $\g(t) = \exp(tv)$. If $S$ is compact and acausal, then $s$ is continuous.  
\end{cor}

\proof
By Propositions \ref{USC s prop} and \ref{LSC s prop}, it suffices to show $s(v) > 0$ for all $v \in \mc{N}S^+$. Suppose, to the contrary, $s(v) = 0$ for some $v \in \mc{N}S^+$. Since the normal exponential map is a local diffeomorphism, there is a neighborhood $U$ of $\g(0)$ such that $\g$ is the unique geodesic from $S$ to any point on $\g$ within $U$.
By strong causality, there is neighborhood $V \subset U$ about $\g(0)$ such that if $\lambda$ is a causal curve with endpoints in $V$, then $\lambda \subset U$. If $s(v) = 0$, then there would be a sequence of causal geodesics $\s_n\colon [0, b_n] \to M$ from $S$ to $\g(1/n)$. By taking a subsequence, we can assume $\s_n(0) \to p$ for some $p \in S$. We have $p \neq \g(0)$ since strong causality implies $\s_n(0) \notin V$. Therefore a limit curve argument \cite[Cor. 3.32]{BEE} implies the existence of a causal curve from $p$ to $\g(0)$, but this contradicts acausality of $S$. 
\qed

\medskip

\subsection{Compactness results - the no horizon case}\label{cpt no horizon sec}

A key ingredient in the proofs of our spacetime topology results is to establish the compactness of  certain regions within a given spacelike hypersurface (specifically, the set  $E_1$ described below).  While our strategy here is similar to some other works (e.g. \cite{Gannon, Lee, Costa, GalBrowdy, LingLesourd}), our approach emphasizes the codimension 2 null cut locus.

Let $(M,g)$ be a spacetime. Let $V$ be an acausal connected spacelike hypersurface. Let $S \subset V$ be a compact connected separating codimension 2 surface. Then $V \setminus S$ is disconnected. Let $E_1'$ and $E_2'$ form a separation for $V \setminus S$. Let $E_1 = E_1' \cup S$ and $E_2 = E_2' \cup S$. Then $E_1$ and $E_2$ are closed sets. Moreover, $V = E_1 \cup E_2$ with $S = E_1 \cap E_2 = \pd_V E_1 = \pd_V E_2$. Connectedness of $E_1$ and $E_2$ follows from connectedness of $V$ and $S$. Let $n$ denote the unit future directed timelike normal vector field on $V$. Let $\nu$ denote the outward unit spacelike normal vector field on $S$ where `outward' means $\nu$ points into the direction of $E_2$. Physically, one should think of $S$ as a surface near `infinity.'

For each $x \in S$, set $v_x = n_x - \nu_x$ and define $\g_x(t) = \exp_x(tv_x)$. Define
\[
t_x \,=\, s(v_x)\,=\, \sup \{t \geq 0 \mid d\big(S, \g_x(t)\big) = 0\}.
\]
Note that if $\g_x$ is defined at $t_x$, then $\g_x(t_x)$ is the $S$-cut point along $\g_x$.

\medskip

\begin{Def}\label{W def}
\emph{We define 
\[
W \,=\, \bigcup_{x \in S} \g_x(I_x)
\]
where $I_x = [0, t_x]$ if $\g_x$ is defined at $t_x$ and $I_x = [0, t_x)$ if $\g_x$ is not defined at $t_x$. Note that $W$ is an achronal set.}
\end{Def}

\medskip

\begin{lem}\label{W cpt lem}
Assume that for each $x \in S$, there exists an $S$-cut point along $\g_x$. Then $W$ is compact.
\end{lem}

\proof
We will show sequential compactness of $W$. Let $q_n$ be any sequence in $W$. There are $t_n \in [0, t_{p_n}]$ such that $q_n = \g_{p_n}(t_n)$. By restricting to a subsequence, we can assume $p_n \to p \in S$. By upper semicontinuity of the $s$ function (Proposition \ref{USC s prop}), the map $x \mapsto t_x$ is upper semicontinuous on $S$, and hence there is a $t_{\rm max}$ such that $t_n \leq t_{\rm max}$ for all $n$. Therefore, by restricting to a further subsequence, we can assume $t_n \to t$. Since $t_n \leq t_{p_n}$, we have $t \leq t_p$ by uppser semicontinuity. Let $q = \g_p(t)$. Then $t \leq t_p$ implies $q \in W$. That $q_n \to q$ follows by continuity of the exponential map.
\qed

\medskip
\medskip

Following Costa e Silva \cite{Costa}, we say that a spacetime $(M,g)$ admits a \emph{piercing} for $V$ if there is a future directed timelike vector field $X$ on $M$ such that each maximal integral curve of $X$ intersects $V$ at exactly one parameter value. We call $X$ a \emph{piercing vector field} for $V$. A spacetime is \emph{past reflecting} if ``$I^+(q) \subset I^+(p)$ implies $I^-(p) \subset I^-(q)$ for all $p$ and $q$" which is equivalent to ``$q \in \ov{J^+(p)}$ implies $p \in \ov{J^-(q)}$ for all $p$ and $q$." See \cite{MinguzziLivRev}.  Note that past reflectivity is implied by closure of $J^+(p)$ for all $p$, which holds, for example, in globally hyperbolic spacetimes.

\medskip
\medskip

\begin{prop}\label{W chara prop}
Suppose $(M,g)$ is past reflecting and admits a piercing for $V$. Then
\[
W \,=\,\pd I^+(E_2) \setminus {\rm int}_V E_2.
\] 
\end{prop}

\medskip

\noindent\emph{Remark.} 
 Any future directed timelike vector field in a  globally hyperbolic spacetime is a piercing for any of its Cauchy surfaces.  Therefore Proposition \ref{W chara prop} holds for globally hyperbolic spacetimes.  

\medskip

\proof
Let $X$ be a piercing vector field for $V$ normalized to $h(X,X) = 1$ where  $h$ is a background complete Riemannian metric on $M$. Since maximal integral curves are inextendible as continuous curves, the integral curves of $X$ have domain $\R$. Let $\phi \colon \R \times M \to M$ denote the flow of $X$. Let $\phi_V \colon \R \times V \to M$ denote the restriction of $\phi$ to $\R \times V$. The piercing assumption and the fact that integral curves don't intersect imply that $\phi_V$ is bijective and hence is a diffeomorphism \cite[Thm. 9.20]{Lee_smooth}.  Let $\pi \colon \R \times V \to V$ denote the natural projection onto $V$. Put $r = \pi \circ \phi_V^{-1}$. Then $r \colon M \to V$ is a smooth retraction of $M$ onto $V$. Lastly, let $\mathcal{C}$ denote the timelike cylinder formed by the image of the integral curves of $X$ through $S$. That is,
\[
\mathcal{C} \,=\,r^{-1}(S) \,=\, \{\phi(t,x) \mid t \in \R \text{ and } x \in S\}. 
\]

Set $T = \pd I^+(E_2) \setminus \text{int}_V E_2$. We want to show $T = W$.

We first show $W \subset T$. Fix $q \in W$. Then $q = \g_x(t)$ for some $t \in [0, t_x]$. If $t = 0$, then $q \in S$ and so acausality of $V$ implies $q \in T$. Now suppose $t > 0$. Then $q \notin \text{int}_V E_2$ since $V$ is acausal; hence it suffices to show $q \notin I^+(E_2)$. Suppose $q \in I^+(E_2)$. We will obtain a contradiction by showing $q \in I^+(S)$. 
Let $\l\colon [0,b] \to M$ denote a timelike curve from $E_2$ to $q$. If $\l(0) \in S$, then we're done, so we can assume $\l(0) \notin S$. Either $r(q) \in E_1$ or $r(q) \in E_2$. First assume $r(q) \in E_1$. Then $r \circ \lambda$ is a path in $V$ that begins in $E_2$ and ends in $E_1$. Since $S$ separates, there is a $t_* \in (0,b]$ such that $r \circ \l(t_*) \in S$. Thus $\l(t_*) \in \mathcal{C}$. Therefore $q \in I^+(S)$ follows by concatenating the integral curve on $\mathcal{C}$ from $S$ to $\l(t_*)$ with the rest of $\l$. Now suppose $r(q) \in E_2$. Since $r \circ \g_x(t') \in E_1$ for sufficiently small $t'$, 
there is a $t_* \in (0,t_x)$ such that $r \circ \g_x(t_*) \in S$. Hence $\g_x(t_*) \in \mathcal{C}$ which implies $q \in I^+(S)$. 

Next we show $T \subset W$.  Define $\mc{H}^+$ to be those points in $J^+(S) \setminus I^+(S)$ which can be reached by $\g_x$ for some $x \in S$. Using past reflectivity, it follows from a key argument in the proof of Theorem 3.5 in \cite{CostaMinguzzi} that $T \subset \mc{H}^+$. Therefore $T \subset J^+(S) \setminus I^+(S)$. Fix $q \in T$.
If $q \in S$, then $q = \g_q(0)$ and so $q \in W$. If $q \notin S$, then there is a null geodesic $\g_x$ such that $q = \g_x(t)$ for some $t > 0$. It suffices to show $t \leq t_x$. If $t_ x = \infty$, then we're done. If $t_x < \infty$ and $t > t_x$, then there is a timelike curve from $S$ to $q$, which is a contradiction.
\qed

\medskip

\begin{prop}\label{W and E cpt prop}
Suppose $(M,g)$ is past reflecting and admits a piercing for $V$. Also, suppose that for each $x \in S$, there is an $S$-cut point along $\g_x$. Then $E_1$ is compact. 
\end{prop}

\proof

Let $r \colon M \to V$ denote the retraction from $M$ to $V$ from the proof of Proposition \ref{W chara prop}. Let $W' = W \setminus S$ and $E_1' = E_1 \setminus S$. Let $\tau = r|_W$ and $\tau' = r|_{W'}$ denote the restrictions of $r$ to $W$ and $W'$. From Proposition~\ref{W chara prop} we have $\tau(W) \subset E_1$. Moreover, $\tau$ is injective since integral curves don't intersect. Since $W'$ is a topological hypersurface, Brower's invariance of domain theorem implies that $\tau'$ is an open map. Since $\tau$ is just the identity on $S$, it follows that $\tau$ is an open map as well. Therefore $\tau(W)$ is open in $E_1$. Since $W$ is compact by Lemma \ref{W cpt lem}, it follows that $\tau(W)$ is closed in $E_1$. Thus $\tau(W) = E_1$ by connectedness of $E_1$ and hence $E_1$ is compact. 
\qed

\medskip
\medskip

In applications to spacetime topology, it is important to identify which notions on a spacetime manifold lift to covers (cf. \cite{CostaMinguzzi}).   An $S$-focal point, as defined in \cite[Def. 10.29]{ON}, will lift to covers by \cite[Prop. 10.30]{ON}.  $S$-cut points, on the other hand, don't necessarily lift to covers as illustrated in the example below. However, by \cite[Prop. 10.48]{ON}, an $S$-focal point implies the existence of an $S$-cut point, which is frequently used in our proofs.

 Let  $S^2$ be the standard round sphere, 
and let $V = \bbR \times S^2$, 
with standard product metric $h$.  
By identifying each of the points $(s,p) \in \bbR \times S^2$ with $(-s,-p) \in  \bbR \times S^2$, we obtain the manifold 
$V'$  which is diffeomorphic to $RP^3$ minus a point, and which inherits a metric $h'$ from $h$. The sphere at $s = 0$ now corresponds to an $RP^2$.  Now consider the spacetime $(M',g') = (\bbR \times V', -dt^2 \oplus h')$, and let $S'$ be a sphere in $V' \equiv \{0\} \times V'$ `parallel' to the $RP^2$.  Each inward  future directed null normal geodesic to $S'$  comes back to $S'$ in the future, which implies that there is an $S'$-cut point along each inward future directed null normal geodesic to $S'$.  Now consider the universal cover $(M,g) = (\bbR \times V, -dt^2 \oplus h)$.
Let $S$ be one of the two $2$-spheres in $V \equiv \{0\} \times V$ covering $S'$.  Now the  inward future directed null normal geodesics to $S$ never return to $S$, and there is no $S$-cut point along any inward future directed null normal geodesic to $S$.  All cut points ``disappear" in the cover.

Proposition \ref{W and E cpt prop} along with \cite[Prop. 10.48]{ON} imply the following corollary.

\medskip

\begin{cor}\label{E1 cpt cor}
Suppose $(M,g)$ is past reflecting and admits a piercing for $V$.  Also, suppose that for each $x \in S$, there is an $S$-focal point along $\g_x$. Then $E_1$ is compact. 
\end{cor}

\medskip

\noindent\emph{Remark.}  Typical ways one ensures the existence of an $S$-focal point is through conditions on the (inward) null expansion on $S$.  The most basic situation is the following.  If we assume (i) the null energy condition holds, $\ric(X,X) \ge 0$ for all null vector $X$, and (ii) $S$ has negative inward null expansion, $\th_- := \div_S v < 0$ (with $v$ as in the beginning of Section \ref{cpt no horizon sec}) then there is an $S$-focal point along each future complete 
$\g_x$.   More generally, there exists an $S$-focal point along each future complete 
$\g_x$, provided (see \cite[Prop. 2]{GalSen}),
\beq \label{focal criterion}
\int_0^{\infty} \ric(\g_x',\g_x')\, ds   > \th_-(x)  \,.
\eeq

\medskip
As a simple application of these ideas  we consider the following generalization of \cite[Thm. 1]{Lee} (see also \cite{Gannon}).
\medskip

\begin{prop} Suppose $(M,g)$ is past reflecting and admits a piercing for $V$.
Let $\mc{C}$ be the `timelike cylinder' formed by the union of the integral curves from $S$ generated by the piercing vector field. If for all $x \in S$, either $\g_x$ meets $\mc{C} \setminus S$ or is future complete and satisfies  \eqref{focal criterion}, then $E_1$ is compact.
\end{prop}

\medskip

\proof
If $\g_x$ meets $\mc{C}\setminus S$, then there is a $t_x < \infty$ such that $\g_x(t_x)$ is a future  $S$-cut point of $S$ along $\g_x$. Suppose $\g_x$ does not meet $\mc{C} \setminus S$. Then $\g_x$ is future complete, and hence by \eqref{focal criterion}, there is a $t$ such that $\g_x(t)$ is an $S$-focal point along $\g_x$. Therefore $\g$ cannot be maximizing past $\g_x(t)$ by \cite[Prop. 10.48]{ON} and hence $t_x < t$. Then compactness of $E_1$ follows by Proposition \ref{W and E cpt prop}.\qed

\subsection{Compactness results - the horizon case}\label{compactness horizon sec}

In this section we extend the main compactness results of the previous section to allow for the presence of horizons (as described below) so as to model the presence of black holes.   This will eventually lead to results about the topology of certain regions outside of these black holes.

Through out this section, we consider a spacetime $(M,g)$  with the following four properties. 

\begin{itemize}

\item[(1)] There is an acausal connected spacelike hypersurface $V$.

\item[(2)] There exists a  codimension 2 surface $\S \subset V$ which separates $V$. ($\S$~may have multiple components.) Let $B'$ and $E'$ form a separation for $V \setminus \S$. Set $B = B' \sqcup \Sigma$ and $E = E' \sqcup \Sigma$.  Not that $\S$ is closed and hence $B$ and $E$ are closed.  Then
\[
V \,=\, B \cup E \quad \text{ and } \quad \Sigma = B \cap E.
\]
We further assume that $B \approx \S \times [0,\e)$. ($V$ only slightly enters the black hole region.)

\item[(3)] There exists a compact connected codimension $2$ surface $S \subset V$ which separates $E \setminus \S$. Let $E_1'$ and $E_2'$ form a separation for $E \setminus (\Sigma \sqcup S)$. Set $E_1 = E_1' \sqcup S \sqcup \S$ and $E_2 = E_2' \sqcup S$. Hence 
\[
E \,=\, E_1 \cup E_2 \quad \text{ and } \quad S \,=\, E_1 \cap E_2.
\]
Connectedness of $E_1$ and $E_2$ follows from connectedness of $V$ and $S$.

\item[(4)] Define $H:= \pd I^+(B) \setminus \text{int}_VB$.  We assume that the null generators of $H$ never leave $H$ when made future inextendible.
\end{itemize}
 
\medskip

\noindent\emph{Remarks.} 

\begin{itemize}

\item[-] When interpreting this model physically, one should view $\S$ as the surface of a black hole, and $S$ as a surface surrounding $\S$.

\item[-] The assumption that the null generators of $H$ never leave $H$ holds whenever $H$ makes up part of an event horizon in the traditional sense (i.e. whenever $H = \big( \pd I^-(\mathcal{J}) \cap M\big) \cap J^+(V)$ where $\mathcal{J}$ is a conformal boundary).

\end{itemize}

\medskip
\medskip

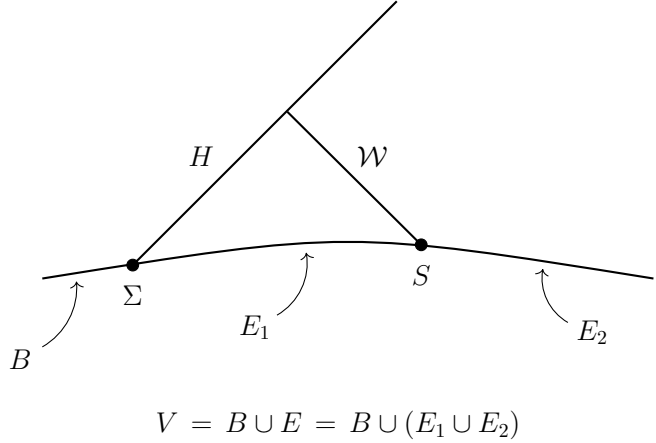
\begin{figure}[t]\label{fig}
\begin{center}
\begin{tikzpicture}[scale = 0.65]

\draw[-, thick] (-6.25,3) .. controls (0,4) .. (6.25,3);

\node [scale = .4] [circle, draw, fill = black] at (-4.4,3.275)  {};
\node(x1) at (-4.4,2.65) [scale = .75] {\large $\Sigma$};

\draw[-, thick] (-4.4,3.275) -- (1,8.675);

\node [scale = .4] [circle, draw, fill = black] at (1.5,3.69)  {};
\node(x2) at (1.5,3.065) [scale = .75] {\large $S$};

\draw[-, thick] (1.5,3.69) -- (-1.2375,6.4125);

\draw [->] (-6.25,1.6) arc [start angle=-60, end angle=5, radius=40pt];
\node at (-6.7,1.35) [scale = .75] {\large $B$};

\draw [->] (-1.45,2.25) arc [start angle=-60, end angle=10, radius=35pt];
\node at (-1.9,2.00) [scale = .75] {\large $E_1$};

\draw [->] (4.5,2.1) arc [start angle=-120, end angle=-190, radius=30pt];
\node at (5,1.9) [scale = .75] {\large $E_2$};

\node at (-3,5.5) [scale = .75] {\large $H$};

\node at (.5,5.5) [scale = .75] {\large $\mc{W}$};

\node(x6) at (-0.2,0) [scale = .75] {\large $V\,=\,B \cup E\,=\,  B \cup \left( E_{1} \cup E_{2}\right) $};

\end{tikzpicture}
\end{center}
\caption{\small{The set up for this section. The black hole horizon is represented by $H = \pd I^+(B) \setminus \text{int}_VB$. }}
\label{set-up fig}
\end{figure}

\medskip
\medskip

As in Definition \ref{W def}, set $W = \bigcup_{x \in S} \g_x(I_x)$. 
Also as in the previous section, we define the upper semicontinuous function
\[
x \mapsto t_x \,=\, s(v_x)\,=\, \sup \{t \geq 0 \mid d\big(S, \g_x(t)\big) = 0\},
\]
so that $\g_x(t_x)$ is an $S$-cut point along $\g_x$ provided $\g_x$ is defined at $t_x$. Under the assumptions of past reflecting and a piercing, we have that $W$ is an achronal topological hypersurface with boundary via the proof of Proposition \ref{W chara prop}. This will be used in the proof of Proposition \ref{S cut cpt horizon prop}.

\smallskip
Henceforth we assume that for  each $x \in S$ either $\g_x$ crosses $H$ or there exists an $S$-cut point along $\g_x$. When we say $\g_x$ \emph{crosses} $H$, we mean that $\g_x$ intersects one of the null generators of $H$ (and hence does so transversely).

 Define $S_H = \{x \in S \mid \g_x \cap H \neq \emptyset\}$. A priori $S_H$ could be the empty set, but under our considerations it will be nonempty. 
 
 \medskip
 
\begin{Def}\label{W hor def}
\emph{Let $\mc{W} \subset W$ be given by}
\[
\mathcal{W} \,=\, \bigcup_{x \in S}\gamma_x\big([0, \tau_x]\big)
\]
\emph{where}
\[
\tau_x \,=\, \left\{
\begin{array}{ll}
      t_x & \text{ if } x \in S^c_H \\
     \min \{t_x, h_x\} & \text{ if } x \in S_H \\
\end{array} ,
\right.
\]
\emph{and where $h_x$ is the unique parameter value such that $\g_x(h_x) \in H$. Here $S_H^c = S \setminus S_H$. Note that $\tau_x < \infty$ for all $x \in S$ by assumption.}
\end{Def}

\medskip

\begin{lem}\label{W cpt lem horizon}
Suppose for each $x \in S$, either $\g_x$ crosses $H$ or there exists an $S$-cut point along $\g_x$. Then $\mc{W}$ is compact and meets $H$. 
\end{lem}

\proof We first show that $\mc{W}$ is compact. Then we show that $S_H \neq \emptyset$ (which means $\mc{W}$ meets $H$). 

A similar argument as used in \cite{GalBrowdy, LingLesourd} shows that $S_H$ is open and hence $S_H^c$ is closed and therefore compact. Since $\pd S_H \subset S_H^c$, it follows from assumption that $\gamma_x(t_x)$ is an $S$-cut point along $\gamma_x$ for each $x \in \pd S_H$. By continuity of the exponential map, for each $x \in \pd S_H$, there is an $\e > 0$ and a neighborhood $U_x \subset S$ of $x$ such that $\g_y$ is defined on $[0, t_x + 2\e]$ for all $y \in U_x$. By upper semicontinuity of $x \mapsto t_x$, we can assume that the neighborhood $U_x$ also satisfies the following property: for each $y \in \ov{U}_x$, we have $t_y < t_x + \e$. Let $\{U_1, \dotsc, U_N\} \subset \{U_x\}_{x \in \pd S_H}$ be a finite subcover of $\pd S_H$. Set $A = S_H^c \cup \ov{U}_1 \cup \dotsb \cup \ov{U}_N$ and $A^c = S \setminus A$. Note that $A$ is compact and $\ov{A^c} \subset S_H$ is a proper subset. Moreover, note that $\g_x(t_x)$ is defined for all $x \in A$ by construction.

We now prove sequential compactness of $\mc{W}$. Let $q_n$ be any sequence in $\mc{W}$. There are $t_n \in [0, \tau_{p_n}]$ such that $q_n = \g_{p_n}(t_n)$. By restricting to a subsequence, we can assume $p_n \to p \in S$. Either infinitely many points of $p_n$ lie in $A$ or infinitely many points of $p_n$ lie in $A^c$. So, by restricting to a subsequence, there are two cases: either the sequence $p_n$ lies in $A$ or the sequence $p_n$ lies in $A^c$. 

Assume the sequence $p_n$ lies in $A$. In this case, we have $t_n \leq \tau_{p_n} \leq t_{p_n} < t_* + \e$ for all $n$. Thus, by restricting to a subsequence, we can assume $t_n \to t \leq t_* + \e$. By upper semicontinuity, we have $t \leq t_p$. Since $A$ is compact, we have $p \in A$ and so $\g_p(t_p)$ is defined and hence $\g_p(t)$ is defined. Set $q = \g_p(t)$. If $p \in S_H^c$, then $\tau_p = t_p$; hence $t \leq \tau_p$, and so $q \in \mc{W}$. If $p \notin S_H^c$, it suffices to show that $t \leq h_p$.  If $t > h_p$, then $q  \in I^+(B)$ (since $\g_p$ intersects $H$ transversely) which implies infinitely many $q_n \in I^+(B)$ which is a contradiction.

Assume the sequence $p_n$ lies in $A^c$. As shown in \cite{GalBrowdy, LingLesourd}, the function $x \mapsto h_x$ is continuous. Since $\ov{A^c} \subset S_H$, it follows that $\sup \{h_{p_n}\} < \infty$. Since $t_n \leq \tau_{p_n} \leq h_{p_n}$, it follows that $t_n \to t$ when restricted to a subsequence. Moreover, taking $n \to \infty$, we have $t \leq h_p$. Therefore $q :=\g_p(t)$ is defined. Upper semicontinuity implies $t \leq t_p$. Thus $t \leq \tau_p$ and so $q \in \mc{W}$.

We now show $S_H \neq \emptyset$. Since $H$ is an achronal boundary, it separates $I^+(V)$ into a future and past set $F = I^+(B)$ and $P = I^+(V) \setminus (F \sqcup H)$ \cite{Penrose}. Seeking a contradiction, suppose $\mc{W}$ does not meet $H$. Then $\mc{W}' \subset P$ where $\mc{W}' = \mc{W} \setminus S$. Let $r \colon M \to V$ denote the flow map of the piercing of $V$ as in Proposition \ref{W chara prop}. By properties of achronal boundaries, we have $r(P) \subset E \setminus \S$ and hence $r(\mc{W}) \subset E \setminus \S$. By Proposition \ref{W chara prop}, we have $\mc{W}$ is a compact achronal $C^0$ hypersurface with $\text{edge}(\mc{W}) = S$; the contradiction now follows as in  \cite{GalBrowdy, LingLesourd} since $\text{edge}(E_1) = \Sigma \cup S$.
\qed

\medskip
\medskip

Analogous to Proposition \ref{W and E cpt prop} and Corollary \ref{E1 cpt cor}, we have the following  proposition and corollary.
\medskip

\begin{prop}\label{S cut cpt horizon prop}
Assume $(M,g)$ is past reflecting and admits a piercing for $V$. Suppose that for each $x \in S$, either $\g_x$ crosses $H$ or there exists an $S$-cut point along $\g_x$. If $H \subset J^+(\S) \setminus I^+(\S)$, then $E_1$ is compact. 
\end{prop}

\proof
The proof follows from similar arguments as in \cite{GalBrowdy, LingLesourd} which we sketch. Let $H_0$ denote the union of the components of $H$ which meet $\mc{W}$. Note that $H_0$ is nonempty by Lemma \ref{W cpt lem horizon}. 
Using the fact that the null generators of $H$ never leave $H$ when made future inextendible, it follows that $H_0$ is a smooth null hypersurface with boundary $\S_0$ (here we have used $H \subset J^+(\Sigma) \setminus I ^+(\Sigma)$), which consists of connected components of $\S$.
Moreover, since the null generators of $H_0$ meet $W$ transversely, it can be shown by an open and closed argument as in \cite{GalBrowdy, LingLesourd} that each null generator of $H_0$ issuing from $\S_0$ meets $\mc{W}$ exactly once.

Let $H_0' \subset H_0$ denote the portion of $H_0$ comprising of the null geodesic generators of $H_0$ which terminate when they intersect $\mc{W}$. As suggested in figure \ref{fig}, $H_0' \cup \mc{W}$ forms a compact achronal topological hypersurface with boundary $\S_0 \cup S$; we now sketch a proof of this. Compactness follows from compactness of $\mc{W}$ (Lemma \ref{W cpt lem horizon}) and compactness of $H_0'$; the latter follows from an analogous argument as in \cite{GalBrowdy, LingLesourd} which uses compactness of the intersection $H_0' \cap \mc{W}$ which implies that $\Sigma_0$ is compact. Using causal theoretic techniques, it's easily shown that $H_0' \cup \mc{W}$ is achronal. Lastly, that $H_0' \cup \mc{W}$ is a topological hypersurface with boundary follows from recognizing that $H_0$ and $W$ are topological hypersurfaces with boundary, and hence away from the boundary each can be locally modeled by graphing functions as in \cite[Lem. 14.25]{ON}; then, for each point in $H_0' \cap \mc{W}$, there is a neighborhood in $H_0' \cup \mc{W}$ that agrees with the set obtained by taking the minimum of the graphing functions associated with $H_0$ and $W$. 

The integral curves of the piercing vector field must map points of $H_0' \cup \mc{W}$ into $E_1$.  It then follows from a connectedness argument using invariance of domain that the piercing vector field maps $H_0' \cup \mc{W}$ onto $E_1$; hence $E_1$ is compact. 
\qed

\medskip

\begin{cor}\label{S focal cpt horizon cor}
Assume $(M,g)$ is past reflecting and admits a piercing for $V$.
Suppose that for each $x \in S$, either $\g_x$ crosses $H$ or there exists an $S$-focal point along $\g_x$. If $H \subset J^+(\Sigma) \setminus I ^+(\Sigma)$, then $E_1$ is compact. 
\end{cor}

\medskip

The following gives a sufficient condition for when $H \subset J^+(\Sigma) \setminus I ^+(\Sigma)$. This will be used in the proofs of our topological censorship results in section  \ref{TopologyHorizon}.

\medskip

\begin{prop}\label{hor geo lem}
Assume $(M,g)$ is past reflecting and admits a piercing for $V$. If $\Sigma$ is compact, then $H \subset J^+(\S) \setminus I^+(\S)$.
\end{prop}

\proof  This follows from an analogous argument as in the proof of Proposition \ref{W chara prop} (specifically, the part where it is shown that $T \subset W$).   The essential part of that proof is based on the proof of Theorem~3.5 in \cite{CostaMinguzzi}.\qed

\medskip

\section{Applications to spacetime topology - the no horizon case}\label{TopologyNoHorizon}

The results we present on spacetime topology, in  both the horizon and no horizon cases, use a proof strategy similar to a number results in the literature; cf. \cite{Gannon, Lee, Costa, CostaMinguzzi, Stein, GalBrowdy, LingLesourd}.   One of our main results, first considered by Costa e Silva~\cite{Costa} in the no horizon setting (see also \cite{CostaMinguzzi}), is to allow $S$ (as in the beginning of  Section \ref{cpt no horizon sec}) to have nontrivial fundamental group.  An essential step in the proof is to construct a certain covering spacetime.  This is discussed in the following section in the no horizon case.  As mentioned in the introduction, our construction 
addresses a certain issue with the covering construction in \cite{Costa}.

\medskip

\subsection{Gluing constructions}\label{gluing sec}
In this section we describe the gluing constructions in the no horizon case.

Let $(M,g)$ be a spacetime. Let $V$ be an acausal connected spacelike hypersurface. Suppose $(M,g)$ admits a piercing for $V$. Let $S \subset V$ be a compact connected separating codimension 2 surface. Then $V \setminus S$ is disconnected. Let $E_1'$ and $E_2'$ form a separation for $V \setminus S$. Let $E_1 = E_1' \cup S$ and $E_2 = E_2' \cup S$. Then $E_1$ and $E_2$ are closed sets. Moreover, $V = E_1 \cup E_2$ with $S = E_1 \cap E_2 = \pd_V E_1 = \pd_V E_2$. Connectedness of $E_1$ and $E_2$ follows from connectedness of $V$ and $S$. 

Let 
\[
i \colon S \to V\, \:\:\:\: i_1 \colon S \to E_1\, \:\:\:\: i_2 \colon S \to E_2
\]
denote the inclusion maps of $S$ into $V$, $E_1$, and $E_2$. 

Let $p \colon \wt{E}_1 \to E_1$ be a covering of $E_1$ such that $p_*\pi_1(\wt{E}_1, \hat{x}) = i_{1*} \pi_1(S, x)$ with $x \in S$ and $\hat{x} \in \wt{S} := p^{-1}(S)$ (such a covering exists by \cite[Cor. 12.19]{Lee_top}). In the general relativity literature, $p$ is known as the \emph{Hawking cover}. Slightly abusing notation, we wish to construct a cover $p \colon \wt{V} \to V$ which extends the covering $p \colon \wt{E}_1 \to E_1$ by gluing appropriate coverings of $E_2$ to each connected component of $\wt{S}$ along the boundary of the covering of $E_2$. This covering will be used to establish spacetime topology results. The next two propositions will aid us in the construction of $\wt{V}$.

\medskip

\begin{prop}\label{covering prop 1}\,
\begin{itemize}
\item[\emph{(1)}] If $\hat{S}$ is the connected component of $\wt{S}$ containing $\hat{x}$, then $\hat{S}$ is isometric to $S$.
\item[\emph{(2)}] If $\wt{S}$ is connected, then $i_{1*} \colon \pi_1(S,x) \to \pi_1(E_1,x)$ is onto.
\end{itemize}
\end{prop}

\proof
We first prove (1). The restriction $p|_{\hat{S}}\colon \hat{S} \to S$ is a surjective local isometry; hence it suffices to show that it's injective. Fix $\hat{y}, \hat{z} \in \hat{S}$ such that $p(\hat{y}) = p (\hat{z})= y \in S$. Let $\hat{\a} \subset \hat{S}$ be a curve connecting $\hat{y}$ to $\hat{x}$ to $\hat{z}$. Let $\a = p \circ \hat{\a}$. Then $\a \subset S$ is a loop based at $y$ which passes through $x$ and hence we can assume $\a$ is based at $x$. Since $p_*\pi_1(\wt{E}_1, \hat{x}) = i_{1*} \pi_1(S,x)$, there is a loop $\wt{\b} \subset \wt{E}_1$ based at $\hat{x}$ such that $\b = p \circ \wt{\b}$ is homotopic to $\a$. Since homotopies lift, $\wt{\b}$ is homotopic to the curve formed by joining $\hat{x}$ to $\hat{z}$ and $\hat{y}$ to $\hat{x}$ via $\hat{\alpha}$. However, the latter is a curve only if $\hat{z} = \hat{y}$.  

Now we prove (2). Assume $\wt{S}$ is connected. Let $\b$ be a loop in $E_1$ based at $x \in S$. Since $\wt{S}$ is isometric to $S$ by (1), the lift of $\wt{\b}$ is a loop based at $\wt{x} = p^{-1}(x) \in \wt{S}$. Then, by hypothesis, there is a loop $\a \subset S$ based at $x$ such that $\beta$ is homotopic to $\a$. \qed

\medskip
\medskip

\noindent\emph{Remarks.}

\begin{itemize}

\item[-] Haggman et al.\ \cite{Haggman} observe (1), based on a common path construction technique for defining the covering $p \colon \wt{E}_1 \to E_1$.
As examples  show (see e.g. \cite[Ex. 2.1]{Haggman}), other components of $\wt{S} = p^{-1}(S)$ need not be isometric to $S$.  This is only assured for the component of $p^{-1}(S)$ containing the base point; cf. also the proof of \cite[Prop. 14.48]{ON}, where the connected component containing the base point is used. This rather  subtle point appears to have been overlooked in the covering construction in \cite{Costa}.

\item[-] There is a situation in which each connected component of $\wt{S}$ is isometric to $S$. Specifically, if $p \colon \wt{E}_1 \to E_1$ is a normal covering, i.e. if $p_*\pi_1(\wt{E}_1, \hat{x})$ is a normal subgroup of $\pi_1(E_1, x)$, it follows from \cite[Thm. 11.34]{Lee_top} that each connected component of $\wt{S}$ is isometric to $S$.

\end{itemize}

\medskip
\medskip

Let $\ov{S}$ be a connected component of $\wt{S}$. Then $p|_{\ov{S}} \colon \ov{S} \to S$ is a covering of $S$. Let $\phi = i_2 \circ p|_{\ov{S}}$ with $\phi(\ov{s}) = x$. Let $q \colon \ov{E}_2 \to E_2$ be a covering of $E_2$ such that $q_*\pi_1(\ov{E}_2,\ov{x}) = \phi_*\pi_1(\ov{S}, \ov{s})$ with $\ov{x} \in \pd \ov{E}_2$ satisfying $q(\ov{x}) = x$. By the general lifting criterion, there is a unique lift $\ov{\phi} \colon \ov{S} \to \ov{E}_2$ of $\phi \colon S \to E_2$ with $\ov{\phi}(\ov{s}) = \ov{x}$ as illustrated in the following diagram:

\begin{center}
\begin{tikzcd}
\ov{S} \arrow[r, "\ov{\phi}"] \arrow[d,"p|_{\ov{S}}"'] \arrow[rd, "\phi"']
& \ov{E}_2 \arrow[d, "q"] \\
S \arrow[r, "i_2"']
&  E_2.
\end{tikzcd}
\end{center}

\begin{prop}\label{covering prop 2}
If $i_{2*}\colon \pi_1(S, x) \to \pi_1(E_2,x)$ is an isomorphism, then $\ov{\phi}$ is an isometry onto $\pd \ov{E}_2$. 
\end{prop}

\proof
We first show $\pd \ov{E}_2$ is connected. Since $\pd \ov{E}_2 = q^{-1}(\pd E_2) = q^{-1}(S)$, it suffices to show $q^{-1}(S)$ is connected. Seeking a contradiction, suppose $\ov{S}_1$ and $\ov{S}_2$ are two disjoint connected components of $q^{-1}(S)$. Let $\ov{x}_1 \in \ov{S}_1$ and $\ov{x}_2 \in \ov{S}_2$ with $q(\ov{x}_1) = q(\ov{x}_2) = x$. Let $\ov{\a}$ be a curve in $\ov{E}_2$ from $\ov{x}_1$ to $\ov{x}_2$. Then $\a = q \circ \ov{\a}$ is a loop in $E_2$ based at $x$. Since $i_{2*}$ is onto, there is a loop $\b$ in $S$ based at $x$ which is homotopic to $\a$. Let $\ov{\b}$ denote the lift of $\b$ starting at $\ov{x}_1$. Since homotopies lift, $\ov{\b}$ is homotopic to $\ov{\a}$ and hence $\ov{\b}$ has endpoint $\ov{x}_2$. But $\ov{\b}$ can't leave $\ov{S}_1$, which is a contradiction.

Set $p' = p|_{\ov{S}}$ and $q' = q|_{\pd \ov{E}_2}$.  Note that $p'$ and $q'$ are covering maps. (We required connectedness of $\pd \ov{E}_2$ here.) Let $\ov{\phi'}\colon \ov{S} \to \pd \ov{E}_2$ be the map satisfying $\ov{\phi} = \ov{i_2} \circ \ov{\phi'}$ where $\ov{i_2}\colon \pd \ov{E}_2 \to \ov{E}_2$ denotes the inclusion map. Then we have the following commutative diagram:
\begin{center}
\begin{tikzcd}[column sep=small]
\ov{S} \arrow[rr, "\ov{\phi'}"]\arrow[dr, "p'"'] & & \pd \ov{E}_2 \arrow[dl, "q'"]
\\
& S.  &
\end{tikzcd}
\end{center}

By the covering isomorphism criterion \cite[Thm. 11.40]{Lee_top}, it suffices to show that $p'_* \pi_1(\ov{S}, \ov{s}) = q_*' \pi_1(\pd \ov{E}_2, \ov{x})$. Since
\begin{align*}
q_* \pi_1(\ov{E}_2, \ov{x}) \,&=\, \phi_* \pi_1(\ov{S}, \ov{s})
\\
\,&=\, i_{2*}\circ p_*' \pi_1(\ov{S}, \ov{s}),
\end{align*}
injectivity of $i_{2*}$ gives 
\[
p_*'\pi_1(\ov{S}, \ov{s}) \,=\, i_{2*}^{-1} \circ q_* \pi_1(\ov{E}_2, \ov{x}).
\]
Therefore it suffices to show
\[
i_{2*}^{-1} \circ q_* \pi_1(\ov{E}_2, \ov{x}) \,=\,  q'_* \pi_1(\pd \ov{E}_2, \ov{x}) .
\]
Both the left and right inclusion of the above equality follow from the following commutative diagram:
\begin{center}
\begin{tikzcd}
\ov{S} \arrow[r, "\ov{\phi'}"]  \arrow[rd, "p'"']
& \pd\ov{E}_2 \arrow[d, "q'"'] \arrow[r, "\ov{i_2}"] & \ov{E}_2 \arrow[d, "q"] \\
&  S \arrow[r, "i_2"'] & E_2.
\end{tikzcd}
\end{center}
To see the right inclusion, fix $[\a] \in q'_*\pi_2(\pd \ov{E}_2, \ov{x})$. Then there is a loop $\ov{\a} \subset \pd \ov{E}_2$ such that $[\a] = [q' \circ \ov{\a}]$. Since $i_{2*}$ is injective, we have $[\a] = i_{2*}^{-1}\big([q \circ \ov{i_2}\circ \ov{\a}] \big)$, which shows the right inclusion.

To see the left inclusion, fix $[\a] \in i_{2*}^{-1} \circ q_* \pi_1(\ov{E}_2, \ov{x})$. Then there is a loop $\ov{\a} \subset \ov{E}_2$ such that $[i_2 \circ \a] = [q \circ \ov{\a}]$. Since $q_*\pi_1\big(\ov{E}_2, \ov{x}\big) = \phi_* \pi_1(\ov{S}, \ov{s})$, there is a loop $\beta \subset \ov{S}$ such that $[q \circ \ov{\a}] = [\phi \circ \b]$. Thus
\[
[\a] \,=\, [i_2^{-1} \circ \phi \circ \b] \,=\, [i_2^{-1} \circ (q \circ \ov{i_2} \circ \ov{\phi'}) \circ \b]\,=\, [q' \circ \ov{\phi'} \circ \b],
\]
which shows the left inclusion.
\qed

\medskip

We can now construct the desired covering $p \colon \wt{V} \to V$. Let $p \colon \wt{E}_1 \to E_1$ be the covering described above Proposition \ref{covering prop 1}. Let $\hat{S}$ be the connected component of $\wt{S}$ containing $\hat{x}$. By Proposition \ref{covering prop 1} (1), $\hat{S}$ is isometric $S$. Therefore, we glue an isometric copy of $E_2$ along $\hat{S}$ in the same way they are glued in the base space $V$. Let $\ov{S}$ be another connected component of $\wt{S}$. Let $q \colon \ov{E}_2 \to E_2$ be the covering described above Proposition \ref{covering prop 2}. Assuming $i_{2*}$ is an isomorphism, Proposition \ref{covering prop 2} implies that we can glue $\ov{E}_2$ along $\ov{S}$. Performing this gluing for each connected component $\ov{S}$ of $\wt{S}$ yields the space $\wt{V}$ along with the covering map $p \colon \wt{V} \to V$ (we're slightly abusing notation here). Note that $p|_{\ov{E}_2} = q$ and $p|_{\wt{E}_1}$ is just the covering we started with. 

Lastly, we note that if $S$ is simply connected, then we do not need to assume that $i_{2*}$ is an isomorphism (although it's automatically injective). This follows because surjectivity of $i_{2*}$ was used in Proposition \ref{covering prop 2} to ensure that $\pd \ov{E}_2$ is connected. In the case that $S$ is simply connected, each connected component $\ov{S}$ of $\wt{S}$ is isometric to $S$ and hence in this case we glue a copy of $E_2$ along $\ov{S}$ in the same way they are glued in the base space $V$; we do not need to assume surjectivity of $i_{2*}$ since $\pd E_{2}$ is automatically connected.  

\medskip

\subsection{Spacetime topology {\`a} l{a} Gannon-Lee}
\label{spacetime top results sec}

\medskip

Let $p \colon \wt{V} \to V$ be the covering constructed in the previous section. We now construct a corresponding spacetime covering $P \colon \wt{M} \to M$ such that $\wt{V}$ embeds into $\wt{M}$ and $P|_{\wt{V}} = p$. 
 We let  $P \colon \wt{M} \to M$ be the covering such that $P_*\pi_1(\wt{M}, \wt{a}) = \psi_*\pi_1(\wt{V},\wt{x})$ where $\psi = \iota \circ p$ and $\iota \colon V \to M$ is the inclusion map.  From the general map lifting criterion, we obtain the following commutative diagram:

\begin{center}
\begin{tikzcd}
\wt{V} \arrow[r, "\ov{\psi}"] \arrow[d,"p"'] \arrow[rd, "\psi"']
& \wt{M} \arrow[d, "P"] \\
V \arrow[r, "\iota"']
&  M.
\end{tikzcd}
\end{center}
Since $\ov{\psi}$ is an embedding\footnote{$\psi$ is an immersion; hence $\ov{\psi}$ is an immersion. Then $\ov{\psi}$ is an embedding follows from $\ov{\psi}$ being an injective open map onto its image. Injectivity of $\ov{\psi}$ follows from the homotopy lifting property applied to the commutative diagram along with the fact that $\iota_*$ is injective which follows from $V$ being a retract of $M$. Then $\ov{\psi}$ is an open map onto its image follows from $p$ and $P$ being local diffeomorphisms along with $\iota$ being a topological embedding.}, we can identify $\wt{V}$ with its image in $\wt{M}$. This completes the construction of the desired spacetime covering. 

Since local geometries are preserved, it follows that $\wt{V}$ is a connected spacelike hypersurface within $\wt{M}$. Moreover, $\wt{V}$ is acausal within $\wt{M}$ since any causal curve connecting two points on $\wt{V}$ would project down to a causal curve in $M$ connecting two points on $V$ which would contradict acausality of $V$ within $M$. Lastly, any piercing vector field $X$ on $M$ for $V$  can be lifted to a piercing vector field $\wt{X}$ on $\wt{M}$ for $\wt{V}$ via the local isometry $P$. That $\wt{X}$ is a piercing vector field for $\wt{V}$ follows from $X$ being one for $V$. 

As in the previous section, let
\[
i \colon S \to V\, \:\:\:\: i_1 \colon S \to E_1\, \:\:\:\: i_2 \colon S \to E_2
\]
denote the inclusion maps of $S$ into $V$, $E_1$, and $E_2$.

\medskip

The following theorem generalizes a well known result of Gannon (\cite[Corollary~1.20]{Gannon}; see also Lee \cite[Theorem 5]{Lee}). Indeed, it is an adaptation of \cite[Theorem~2.1]{Costa} (see also \cite{CostaMinguzzi, Stein}).  
Our modification of
the covering space argument in \cite{Costa} (see the remark after 
Proposition~\ref{covering prop 1})
has required a strengthening of the assumption on $i_{2*}$; see Proposition \ref{covering prop 2}. The assumption of global hyperbolicity used will be weakened some in section \ref{timelikebdry}.

\medskip

\begin{thm}\label{onto}
Suppose $V$ is a Cauchy surface for $(M,g)$, and suppose there exists an $S$-focal point along $\g_x$ for each $x \in S$. In the case that $S$ is not simply connected, assume $i_{2*}\colon \pi_1(S, x) \to \pi_1(E_2,x)$ is an isomorphism. If $E_2$ is noncompact, then $i_{1*} \colon \pi_1(S,x) \to \pi_1(E_1, x)$ is onto. In particular, $E_1$ is simply connected whenever $S$ is simply connected and $E_2$ is noncompact. 
\end{thm}

It follows from Seifert-Van Kampen that  $i_{1*} \colon \pi_1(S,x) \to \pi_1(V, x)$ is onto. 

\proof
Let $p \colon \wt{E}_1 \to E_1$ be the covering of $E_1$ such that $p_*\pi_1(\wt{E}_1, \hat{x}) = i_{1*}\pi_1(S,x)$ with $\hat{x} \in \wt{S} := p^{-1}(S)$. Slightly abusing notation, let $p \colon \wt{V} \to V$ denote the covering constructed at the end of section \ref{gluing sec}. Let $P \colon \wt{M} \to M$ denote the spacetime covering constructed in the beginning of section \ref{spacetime top results sec}. Since $V$ is a Cauchy surface for $M$, it follows that $\wt{V}$ is a Cauchy surface for $\wt{M}$. Hence $(\wt{M},\wt{g})$ is globally hyperbolic and thus is past reflecting and admits a piercing for $\wt{V}$. 

Seeking a contradiction, suppose $i_{1*}$ is not onto. By Proposition \ref{covering prop 1}(2), we have $\wt{S}$ is disconnected. Let $\hat{S}$ denote the connected component of $\wt{S}$ containing $\hat{x}$, and let $\ov{S}$ denote some other connected component. Let $D_2 \subset \wt{V}$ denote the isometric copy of $E_2$ attached to $\hat{S}$. Let $D_1 = (\wt{V} \setminus D_2) \sqcup \hat{S}$. Since focal points lift, we can apply Corollary~\ref{E1 cpt cor} in the covering spacetime $\wt{M}$ to conclude that $D_1$ is compact. However, $D_1$ contains the covering $\ov{E}_2$ of $E_2$ which is attached along $\ov{S}$. Since $E_2$ is noncompact, it follows that $\ov{E}_2$ is noncompact and hence $D_1$ is noncompact, which is a contradiction.
\qed

\medskip
\medskip

Theorem \ref{onto} can be interpreted as an incompleteness theorem in the following way. Assume $E_2$ is noncompact, $i_{2*}$ is an isomorphism, the null energy condition holds, and $S$ has negative inward null expansion.  Then, if $i_{1*}$ is not onto, some inward null geodesic from $S$ must be future incomplete since otherwise there would be an $S$-focal point along each $\g_x$ for each $x \in S$ (see the remark after Corollary \ref{E1 cpt cor}). More generally, if the inward null geodesics are future complete, then  \eqref{focal criterion} must be violated along at least one of the null geodesics.

The following examples illustrate how Theorem \ref{onto} can be interpreted as an incompleteness theorem.

\medskip
\medskip

\noindent\emph{Examples.}

\begin{itemize}

\item[(1)] Consider the Schwarzschild $\R P^3$ geon described in \cite{FSW93}. In this case, $V \approx \R P^3 \setminus\{\text{pt}\}$. Take $S \approx S^2$ to be a surface of constant latitude above the $\R P^2$ equator. $S$ separates $V$. Then $E_1$ corresponds to the ``inside" which includes the $\R P^2$, and let $E_2$ be the ``outside". In this case $S$ is simply connected, but $E_1$ is not simply connected. As in Schwarzschild, $S$ has negative inward null expansion and since the spacetime is vacuum, the null energy condition holds. Hence by Theorem \ref{onto} there must be some future directed null geodesic from $S$ which is incomplete, which, of course, holds for Schwarzschild $\R P^3$ geon. 

\item[(2)] We construct a time-symmetric three-dimensional spatial slice $V$ as follows. Let $\mathcal{V} = (S^1 \times S^2) \# S^3$. Fix a point $P \in \mathcal{V}$, and let $V = \mc{V} \setminus\{P\}$. Since $\mc{V}$ admits a metric of positive scalar curvature, there is a metric $h$ on $V$ with zero scalar curvature such that $V$ is asymptotically flat, see \cite{Schoen_notes, Lee_Parker}, where $P$ represents the point of infinity. We can pick a surface $S \approx S^2$ surrounding $P$ such that $S$ has positive $h$-mean curvature with respect to the outward unit normal (i.e. towards $P$). Then $E_2 \approx S \times[0,\infty)$. A couple applications of the Seifert-Van Kampen theorem show that $\pi_1(E_1) \approx \pi_1(V) \approx \pi_1(\mc{V}) \approx \Z$. Since $(V,h)$ has zero scalar curvature, it satisfies the constraint equations for a time-symmetric initial data slice. Let $(M,g)$ denote the maximal globally hyperbolic development of $(V,h,K = 0)$ for the vacuum Einstein equations. Since $S$ has positive mean curvature and $K = 0$, $S$ is inner trapped within $(M,g)$.  $(M,g)$ is vacuum, so the null energy condition holds. Lastly, $i_{1*}$ is not onto since $S$ is simply connected but $E_1$ is not. Thus, by Theorem \ref{onto}, $(M,g)$ must contain an incomplete future directed null geodesic emanating inwards from $S$. In fact this incompleteness conclusion can be derived from any closed, nonsimply connected manifold $\mc{V}$ with positive scalar curvature in dimensions $\geq 3$.

\item[(3)] In examples (1) and (2), $S$ is topologically a sphere. It is easy to modify example~(2) to show how Theorem \ref{onto} can be used to produce an incompleteness theorem even when $S$ is not simply connected. Let $V$, $S$, $E_1$, and $E_2$ be as in example (2). Let $h$ denote the metric on $V$ with zero scalar curvature. Let $\bar{V} = S^1 \times V$ and $\bar{h} = ds^2 \oplus h$. Then $(\bar{V}, \bar{h})$ has zero scalar curvature as well. Let $\bar{S} = S^1 \times S$. Then $\bar{S}$ separates $\bar{V}$ via $\bar{E}_1 = E_1 \times S^1$ and $\bar{E}_2 = E_2 \times S^1$. The $\bar{h}$-mean curvature of $\bar{S}$ within $\bar{V}$ equals the $h$-mean curvature of $S$ within $V$. Therefore $\bar{S}$ is inner trapped within the maximal globally hyperbolic development of the new initial data set, $(\bar{V}, \bar{h}, 0)$, for the vacuum Einstein equations. Since $\pi_1(\bar{S}) \approx \Z$ and $\pi_1(\bar{E}_1) \approx \Z \times \Z$, it follows that $i_{1*}$ is not onto. Thus, by Theorem \ref{onto}, there is an incomplete future directed null geodesic emanating inwards from $\bar{S}$.

\end{itemize}

\medskip

\subsection{Exterior foliations by spacetimes-with-timelike-boundary}\label{timelikebdry}

\medskip

In this section, we will obtain a version of Theorem \ref{onto} for a class of spacetimes that are not in general globally hyperbolic, but which admit, in a certain sense, a foliation by globally hyperbolic spacetimes-with-timelike-boundary. The Birmingham-Kottler spacetimes \cite{Birmingham} and the Horowitz-Myers soliton \cite{HorowitzMyers} are examples that admit such a foliation. 
The motivation here is to extend our results from the previous section
to a setting applicable to asymptotically locally AdS spacetimes.

Let $(\mathcal{M},\mathfrak{g})$ be a spacetime-with-timelike-boundary; see \cite{SanchezAke} for basic definitions and properties.  Then 
\[
\mathcal{M} \,=\, M \cup \pd \mathcal{M}
\]
where $M = \text{int}(\mathcal{M})$. Set $g = \mathfrak{g}|_M$. Note that $(M,g)$ is a spacetime.

\medskip

\begin{lem}\label{past reflect lem}
Let $(\mathcal{M},\mathfrak{g})$ be a spacetime-with-timelike-boundary. If $(\mathcal{M}, \mathfrak{g})$ is past reflecting, then $(M,g)$ is past reflecting.
\end{lem}

\proof
We abbreviate $I^+(p,M)$ with $I^+_M(p)$. Likewise with $I^+_\mc{M}(p)$. 

Fix $p, q \in M$ such that $I^+_M(q) \subset I^+_M(p)$. We want to show $I^-_M(p) \subset I^-_M(q)$. We first show $I^+_{\mc{M}}(q) \subset I^+_{\mc{M}}(p)$. Fix $x \in I^+_{\mc{M}}(q)$. Either $x \in M$ or $x \in \pd \mc{M}$. If $x \in M$, then 
\[
x \,\in\, I^+_\mc{M}(q) \cap M \,=\, I^+_M(q) \,\subset\, I^+_M(p) \,\subset\, I^+_\mc{M}(p).
\]
The equality follows from \cite[Prop.\ 2.6(d)]{SanchezAke}. Suppose $x \in \pd \mc{M}$. Let $\l \subset \mc{M}$ be a timelike curve from $q$ to $x$. Since $q \in M$ and $M$ is open in $\mathcal{M}$, it follows that $\l$ intersects $M$ at some point $r$. Then $r \in I^+_{\mc{M}}(q) \cap M$ and reasoning as above, we have $r \in I^+_{\mc{M}}(p)$ and hence $x \in I^+_{\mc{M}}(p)$. 

By past reflecting of $\mc{M}$, we have $I^-_{\mc{M}}(p) \subset I^-_{\mc{M}}(q)$. Intersecting both sides with $M$ and applying  \cite[Prop.\ 2.6(d)]{SanchezAke} again, we have $I^-_M(p) \subset I^-_M(q).$
\qed

\medskip

\begin{thm}\label{thm timelike bdy}
Suppose $(M,g)$ is a spacetime with $V$, $S$, $E_1$, and $E_2$ given as in the beginning of section \ref{cpt no horizon sec}. We make the following assumptions.
\begin{itemize}
\item[\emph{(a)}] $X$ is a piercing vector field for $V$.

\item[\emph{(b)}] $E_2 \approx S \times [0, \infty)$. 

\item[\emph{(c)}] Set $E_{2,t} \approx S \times [0,t]$ and $V_t = E_1 \cup E_{2,t}$. Let $M_t$ denote the union of the images of the maximal integral curves of $X$ starting on $V_t$. We assume that $M_t$ is a globally hyperbolic spacetime-with-timelike-boundary with Cauchy surface $V_t$. 

\item[\emph{(d)}] There is a $T > 0$ such that for all $x \in S$, there is an $S$-focal point along $\g_x$ within $M_T'$, where $M_T'$ is the interior of $M_T$. 


\end{itemize}
Then $i_{1*} \colon \pi_1(S,x) \to \pi_1(E_1, x)$ is onto. In particular, $E_1$ is simply connected whenever $S$ is simply connected.

\end{thm}

\medskip

\noindent\emph{Remarks.} 

\begin{itemize}

\item[-] By ``there is an $S$-focal point along $\g_x$ within $M_T'$" we mean that if $\g_x(b)$ is an $S$-focal point of $\g_x$, then $\g_x|_{[0,b]}$ lies within $M_T'$. This terminology will also be used in Theorem \ref{last thm}.

\item[-] It's not hard to see that in the case $S$ is inner trapped, $M$ is future null complete, and satisfies the null energy condition, then assumption (d) holds. One may ask if assumption (d) holds when one merely assumes that there is an $S$-focal point along $\g_x$ for each $x \in S$.

\item[-] Anti-de Sitter space is an obvious example of the theorem. In this case, $S$ is topologically the $(n-1)$-sphere and $E_1$ is topologically the $n$-ball. Hence both $S$ and $E_1$ are simply connected.

\item[-] A less trivial example of the theorem is the Horowitz-Myers soliton \cite{HorowitzMyers}. These class of spacetimes are static solutions of the Einstein equations with a negative cosmological constant. In this case, $S$ is topologically an $(n-1)$-dimensional torus with positive mean curvature, while the topology of $E_1$ is the product of a disc and an $(n-2)$-dimensional torus. Therefore $i_{1*}$ is onto but not an isomorphism.

\end{itemize}

\medskip

\proof
Let $p \colon \wt{E}_1 \to E_1$ be the covering of $E_1$ such that $p_*\pi_1(\wt{E}_1, \hat{x}) = i_{1*}\pi_1(S,x)$ with $\hat{x} \in \wt{S} := p^{-1}(S)$. Slightly abusing notation, let $p \colon \wt{V} \to V$ denote the covering constructed at the end of section \ref{gluing sec}. Let $P \colon \wt{M} \to M$ denote the spacetime covering constructed in the beginning of section \ref{spacetime top results sec}. Let $\hat{S}, \wt{E}, D_1,$ and $D_2$ be given as in Theorem \ref{onto}. Let $\wt{M}_T = P^{-1}(M_T)$ and $\wt{M}_T'$ be its interior. Note that $\wt{M}_T$ is the union of the lift of the maximal integral curves which intersect $V_T$. Since focal points lift, assumption (d) holds in $\wt{M}_T'$ with respect to $\hat{S}$. Hence for each $x \in \hat{S}$, there is an $S$-cut point along $\g_x$ with respect to the spacetime $\wt{M}_T'$. 

 Since $\wt{V}_T = P^{-1}(V_T)$ is a Cauchy surface for $\wt{M}_T$, it follows that $\wt{M}_T$ is a globally hyperbolic spacetime-with-timelike-boundary, and hence $\wt{M}_T$ is past reflecting \cite{SanchezAke}. By Lemma \ref{past reflect lem}, it follows that the interior  $\wt{M}_T'$ of $\wt{M}_T$ is past reflecting. Since $X$ lifts to a piercing $\wt{X}$ on $\wt{M}$ for $\wt{V}$, there is a piercing for the interior $\wt{V}_T'$ of $\wt{V}_T$. Thus $D_1 \cap \wt{M}_T'$ is compact by 
 Corollary \ref{E1 cpt cor}.

If $i_{1*}$ is not onto, then $\wt{S}$ contains a component $\ov{S} \neq \hat{S}$ by Proposition \ref{covering prop 1}(2). Then $D_1 \cap \wt{M}_T'$ contains the noncompact end $\ov{S} \times [0,T)$, which is a contradiction. 
\qed

\medskip

The spacetimes in Theorem \ref{thm timelike bdy} are not  in general causally simple. It would be of interest to determine whether or not these spacetimes and their covers are past reflecting. If a spacetime admits a conformal completion which is a globally hyperbolic spacetime-with-timelike-boundary, then the original spacetime is past reflecting via Lemma \ref{past reflect lem}.

\medskip

\section{Applications to spacetime topology - the horizon case}\label{TopologyHorizon}

\medskip

The topological censorship results in this section generalize aspects of \cite{GalBrowdy, LingLesourd}. For more on topological censorship and references to other related results, see \cite{Chrubook}.

\medskip

\subsection{Gluing constructions}\label{glue hor sec}

We consider the setting of a black hole with a horizon in the spacetime. Let $(M,g)$ satisfy properties (1) - (4) in section \ref{compactness horizon sec}. Define $F_1 = B \cup E_1$ and consider the inclusion maps 
\[
i \colon S \to V\, \:\:\:\: i_1 \colon S \to F_1\, \:\:\:\: i_2 \colon S \to E_2.
\]
Let $p \colon \wt{F}_1 \to F_1$ be a covering of $F_1$ such that $p_* \pi_1 (\wt{F}_1, \hat{x}) = i_{1*}\pi_1(S,x)$ with $x \in S$ and $\hat{x} \in \wt{S} := p^{-1}(S)$. Analogous to Proposition \ref{covering prop 1}, we have

\medskip

\begin{prop}\label{covering prop horizon case}\,
\begin{itemize}
\item[\emph{(1)}] If $\hat{S}$ is the connected component of $\wt{S}$ containing $\hat{x}$, then $\hat{S}$ is isometric to $S$.
\item[\emph{(2)}] If $\wt{S}$ is connected, then $i_{1*} \colon \pi_1(S,x) \to \pi_1(F_1,x)$ is onto.
\end{itemize}
\end{prop}

\medskip

Proposition \ref{covering prop 2} also holds in this setting: if $i_{2*}$ is an isomorphism, then for each connected component $\ov{S} \subset \wt{S}$, there is a covering $q \colon \ov{E}_2 \to E_2$ (defined in the same way as before) such that $\ov{S}$ is isometric to $\pd \ov{E}_2$. 

Using the same construction at the end of section \ref{gluing sec} (and slightly abusing notation again), we have a covering $p \colon \wt{V} \to V$ such that $p|_{\ov{E}_2} = q$ and $p|_{\wt{F}_1}$ is just the covering we started with. As in the beginning of section \ref{spacetime top results sec}, there is a corresponding spacetime covering $P \colon \wt{M} \to M$ such that $\wt{V}$ embeds into $\wt{M}$ and $P|_{\wt{V}} = p$. 

\medskip

\subsection{The globally hyperbolic setting}

\medskip
 
Let $(M,g)$ be a spacetime satisfying properties (1) - (4) in section \ref{compactness horizon sec}. In this section we assume $V$ is a Cauchy surface.
Analogous to Theorem \ref{onto}, we have the following theorem.

\medskip

\begin{thm}\label{onto hor}
Let $(M,g)$ satisfy properties \emph{(1)} - \emph{(4)} in section \emph{\ref{compactness horizon sec}} and assume that $V$ is a Cauchy surface and $\Sigma$ is compact.  Suppose that for each $x \in S$, either $\g_x$ crosses $H$ or there exists an $S$-focal point along $\g_x$. In the case that $S$ is not simply connected, assume $i_{2*}\colon \pi_1(S, x) \to \pi_1(E_2,x)$ is an isomorphism. If $E_2$ is noncompact, then $i_{1*} \colon \pi_1(S,x) \to \pi_1(F_1, x)$ is onto. In particular, $F_1$ (and hence $E_1$) is simply connected whenever $S$ is simply connected and $E_2$ is noncompact. 
\end{thm}

\proof
Let $p: \wt{F}_1 \to F_1$ be the covering of $F_1$ such that $p_*\pi_1(\wt{F}_1, \hat{x}) = i_{1*}\pi_1(S,x)$ with $\hat{x} \in \wt{S}:= p^{-1}(S)$. Let $\wt{\Sigma} = p^{-1}(\Sigma)$ and likewise with $\wt{B}$. Since $B = \S \times [0,\e)$, we have $\wt{B} = \wt{\S} \times [0,\e)$.

Extend this covering to $p \colon \wt{V} \to V$ as constructed at the end of section \ref{glue hor sec}, and let $P \colon \wt{M} \to M$ denote the corresponding spacetime covering. Since $V$ is a Cauchy surface for $M$, it follows that $\wt{V}$ is a Cauchy surface for $\wt{M}$.

Seeking a contradiction, suppose $i_{1*}$ is not onto. Then $\wt{S}$ is disconnected by Proposition \ref{covering prop horizon case}(2). Let $\hat{S}$ denote the connected component of $\wt{S}$ containing $\hat{x}$, and let $\ov{S}$ denote some other connected component. Let $D_2 \subset \wt{V}$ denote the isometric copy of $E_2$ attached to $\hat{S}$. Let $D_1 = (\wt{E} \setminus D_2) \sqcup \hat{S}$ where $\wt{E} = p^{-1}(E)$. Note that $D_1$ is noncompact since it contains a copy of $\ov{E}_2$ attached to some $\ov{S}$.

Set $\mc{H} = \pd I^+(\wt{B}) \setminus \text{int}_{\wt{V}}\wt{B}$. That the null generators of $\mc{H}$ never leave $\mc{H}$ follows from the same property holding for $H$.  Next we show that $\mc{H} \subset J^+(\wt{\Sigma}) \setminus I^+(\wt{\S})$. Note that this step does not follow from Proposition \ref{hor geo lem} since $\wt{\S}$ is not necessarily compact. Fix $\wt{y} \in \mc{H}$ and let $y = P(\wt{y})$. By lifting of curves, it suffices to show that $y \in J^+(\S) \setminus I^+(\Sigma)$. 
Note that $\wt{y} \in \mc{H}$   implies that $y \in \ov{J^+(B)} \setminus I^+(B)$ and $y \notin \text{int}_V B$.
Therefore $y \in H$.\footnote{We just observed that $P(\mc{H}) \subset H$. Similarly, it's not hard to see that $\mc{H} = P^{-1}(H)$; this holds in the next section as well.} Since $\Sigma$ is compact, we have $y \in J^+(\Sigma) \setminus I^+(\Sigma)$ by Proposition  \ref{hor geo lem}.

We wish to apply Corollary \ref{S focal cpt horizon cor} to the covering spacetime $\wt{M}$ with $D_1$ and $\mc{H}$ playing the roles of $E_1$  and $H$ in the statement of Corollary \ref{S focal cpt horizon cor}. Since $\wt{M}$ is globally hyperbolic, it's past reflecting and admits a piercing for $\wt{V}$.

Hence, it remains to show that for each $y \in \hat{S}$, either $\g_{y}$ crosses $\mc{H}$ or there exists an $\hat{S}$-focal point along $\g_{y}$. Suppose there does not exist an $\hat{S}$-focal point along $\g_{y}$. Since focal points lift, there does not exist an $S$-focal point along $P \circ \g_y$. Let $x = p(y) \in S$. Then $\g_x = P \circ \g_y$. By assumption, $\g_x$ must cross 
$H = \pd I^+(B) \setminus \text{int}_V B$, and hence enters  $I^+(B)$. 
Since $I^+(\wt{B}) = P^{-1} \big(I^+(B)\big)$, $\g_y$ must cross $\mc{H}$.
\qed

\medskip
\medskip

\subsection{Exterior foliations by spacetimes-with-timelike-boundary}
\medskip

In this section we obtain a version of Theorem \ref{thm timelike bdy} in the setting of black holes. The Birmingham-Kottler spacetimes \cite{Birmingham} are the main examples of the Theorem \ref{last thm}; for these class of spacetimes, $i_{1*}$ is an isomorphism.

\medskip
\medskip

\begin{thm}\label{last thm}
Suppose $(M,g)$ is a spacetime satisfying properties \emph{(1)} - \emph{(4)} in 
section~\emph{\ref{compactness horizon sec}} and $\Sigma$ is compact. We make the following assumptions.
\begin{itemize}

\item[\emph{(a)}] $(M,g)$ admits a piercing for $V$. 

\item[\emph{(b)}] $E_2 \approx S \times [0, \infty)$.
\item[\emph{(c)}] Set $E_{2,t} = S \times [0,t]$ and $V_t = B \cup E_1 \cup E_{2,t}$. Let $M_t$ be the union of the images of the maximal integral curves of $X$ starting on $V_t$. We assume that $M_t$ is a globally hyperbolic spacetime-with-timelike-boundary with Cauchy surface $V_t$.

\item[\emph{(d)}] There is a $T > 0$ such that the following hold:
\bit
 \item[\emph{(i)}] $I^+(B) \cap M_T' \subset I^+(B, M_T')$, where $M_T'$ is the interior of $M_T$.
 \item[\emph{(ii)}] For each $x \in S$, either $\g_x$ crosses $H$ within $M_T'$ or there is an $S$-focal point along $\g_x$ within $M_T'$.
 \eit
\end{itemize}
Then $i_{1*}\colon \pi_1(S,x) \to \pi_1(F_1,x)$ is onto. In particular, $F_1$ (and hence $E_1$) is simply connected whenever $S$ is simply connected. 
\end{thm}

\medskip

\noindent\emph{Remarks} By ``$\g_x$ crosses $H$ within $M_T'$" we mean that if $\g_x$ crosses $H$ at $\g_x(b)$, then $\g_x|_{[0,b]}$ lies within $M_T'$. See also the remark after Theorem \ref{thm timelike bdy}.

\medskip

\proof
Let $p \colon \wt{F}_1 \to F_1$ be the covering of $F_1$ such that $p_*\pi_1(\wt{F}_1, \hat{x}) = i_{1*}\pi_1(S,x)$ with $\hat{x} \in \wt{S} := p^{-1}(S)$. Extend this covering to $p \colon \wt{V} \to V$ as in the end of section \ref{glue hor sec}. Let $P \colon \wt{M} \to M$ denote the corresponding spacetime covering. Let $\hat{S}$, $\wt{E}$, $\wt{B}$, $D_1$, $D_2$, and $\mc{H}$ be given as in the proof of Theorem \ref{onto hor}. 
Likewise, the null generators of $\mc{H}$ do not leave $\mc{H}$ to the future since this property holds for $H$.

Let $X$ be the piercing vector field for $V$. Let $\wt{X}$ denote the lift of $X$ to $\wt{M}$. Since $\wt{X}$ is a piercing vector field for $\wt{V}$, its integral curves produce a retraction $r \colon \wt{M} \to \wt{V}$. Consider the globally hyperbolic spacetime-with-timelike-boundary $M_T$. Since maximal integral curves lift to maximal integral curves, $\wt{M}_T := P^{-1}(M_T)$ is the union of the maximal integral curves of $\wt{X}$ starting on $\wt{V}_T := P^{-1}(V_T)$. Consequently, properties (i) and (ii) from assumption (d) lift to the cover:
\begin{itemize}
\item[$\wt{(\rm{i})}$] $I^+(\wt{B}) \cap \wt{M}_T' \subset I^+(\wt{B}, \wt{M}_T')$, where $\wt{M}_T'$ is the interior of $\wt{M}_T$. 

\item[$\wt{(\rm{ii})}$] For each $\hat{x} \in \wh{S}$, either $\g_{\hat{x}}$ crosses $\mc{H}$ within $\wt{M}_T'$ or there is an $\wh{S}$-focal point along $\g_{\hat{x}}$ within $\wt{M}_T'$.
\end{itemize}
Set $\mc{H}_T = \mc{H} \cap \wt{M}_T'$. From $\wt{(\rm{i})}$, it follows that  $\mc{H}_T  = \pd I^+(\wt{B}, \wt{M}_T') \setminus \text{int}_{\wt{V}_T'}\wt{B}$. Also, we have
 $\mc{H}_T \subset J^+(\wt{\S}, \wt{M}_T') \setminus I^+(\wt{\S}, \wt{M}_T')$; this follows since the same property holds for $H_T = H \cap M_T'$ which lifts to the cover via the same argument used in Theorem \ref{onto hor}.
Moreover, the null generators of $\mc{H}_T$ never leave $\mc{H}_T$ within $\wt{M}_T'$. 
Noting also that $\wt{M}_T'$ is past reflecting by Lemma \ref{past reflect lem}, it now follows from
Corollary~\ref{S focal cpt horizon cor} that $D_1 \cap \wt{M}_T'$ is compact.

If $i_{1*}$ is not onto, then $\wt{S}$ contains a component $\ov{S} \neq \hat{S}$ by Proposition \ref{covering prop horizon case}(2). Then $D_1 \cap \wt{M}_{T}'$ contains the noncompact end $\ov{S} \times [0, T)$, which is a contradiction.\qed

\medskip
\medskip

\noindent
\textsc{Acknowledgements.}  This work was supported in part by a grant from the Simons Foundation (850541, GJG). Eric Ling gratefully acknowledges being supported by the Harold H. Martin Postdoctoral Fellowship at Rutgers University. The authors would like to thank Ivan Costa e Silva for comments on an earlier draft and useful discussions.

\appendix


\end{document}